\documentclass[letterpaper]{article} 

\usepackage[draft]{aaai2026}
\usepackage{times} 
\usepackage{helvet}  
\usepackage{courier}  
\usepackage[hyphens]{url}
\usepackage{graphicx}  
\usepackage{natbib} 
\usepackage{caption} 
\usepackage{algorithm}
\usepackage{algorithmic}
\usepackage{amsmath} 

\usepackage{newfloat}
\usepackage{listings}
\DeclareCaptionStyle{ruled}{labelfont=normalfont,labelsep=colon,strut=off}   
\floatstyle{ruled}
\newfloat{listing}{tb}{lst}{}
\floatname{listing}{Listing}

\title{Chrysalis: A Unified System for Comparing Active Teaching and Passive Learning with AI Agents in Education}
\author{
    Prashanth Arun \textsuperscript{\rm 1}\textsuperscript{\rm 2},
    Vinita Vader \textsuperscript{\rm 1},
    Erya Xu \textsuperscript{\rm 1},
    Brent McCready-Branch \textsuperscript{\rm 1},
    Sarah Seabrook \textsuperscript{\rm 1},
    Kyle Scholz \textsuperscript{\rm 1},
    Ana Crisan \textsuperscript{\rm 1}, 
    Igor Grossmann \textsuperscript{\rm 1},
    Pascal Poupart \textsuperscript{\rm 1}\textsuperscript{\rm 2},
}
\affiliations{
    \textsuperscript{\rm 1}University of Waterloo\\
    \textsuperscript{\rm 2}Vector Institute for AI\\
    \{parun, v2vader, e2xu, bmccreadybranch, samseabrook, kwscholz, amcrisan, igrossma, ppoupart\}@uwaterloo.ca
}

\begin{document}

\maketitle

\begin{abstract}

AI-assisted learning has seen a remarkable uptick over the last few years, mainly due to the rise in popularity of Large Language Models (LLMs). Their ability to hold long-form, natural language interactions with users makes them excellent resources for exploring school- and university-level topics in a dynamic, active manner. We compare students’ experiences when interacting with an LLM companion in two capacities: tutored learning and learning-by-teaching. We do this using Chrysalis, an LLM-based system that we have designed to support both AI tutors and AI teachable agents for any topic. Through a within-subject exploratory study with $N=36$ participants, we present insights into student preferences between the two strategies and how constructs such as intellectual humility vary between these two interaction modes. To our knowledge, we are the first to conduct a direct comparison study on the effects of using an LLM as a tutor versus as a teachable agent on multiple topics. We hope that our work opens up new avenues for future research in this area.

\end{abstract}

\section{Introduction}

Large Language Models (LLMs) bring many new opportunities to the field of education through their ability to hold open-ended freeform natural language conversations. In this work, we focus on two dimensions of AI in education: intelligent tutoring systems, and learning by teaching. One-on-one tutoring has been shown to be significantly more effective than group learning \cite{bloom-2-sigma:84} and LLMs, given their aforementioned ability to tailor experiences to each user, have been seen as a promising option to address the growing need of personalized learning. Learning-by-teaching is a learning paradigm by which students attempt to teach others content that they themselves are learning. Works like \citet{roscoe-tutor-learning:07} discuss the benefits of peer-tutoring -- for example, by eliciting self-questioning and meta-cognitive responses in the teaching student -- while \citet{chase-protege-effect:08} describe a phenomenon known as the prot\'{e}g\'{e} effect, whereby students are more invested in the material when they have to tutor -- or, more broadly, mentor -- another person. \citet{debbane-lbt-challenges:23} highlight the roadblocks that students face when attempting to teach a person, such as psychological barriers and lack of organization. LLMs are promising alternatives whereby students can conduct a natural language instruction without fear of judgment.

Although there has been much work on AI tutors and AI as a teachable agent, we see a dearth in literature that compares how students respond to these two pedagogical approaches. We believe that comparative studies are needed to better cater to individual learning requirements by identifying traits that predispose students to either approach. We also believe that AI applications should be generalizable to any field for maximum accessibility.

In this work, we introduce Chrysalis, a new LLM-based AI companion designed to help students learn from scratch and deepen their existing understanding of any topic. Our system supports 2 modes: AI tutoring (LLM tutor teaches a student) and learning-by-teaching (the student teaches an LLM agent). We use Chrysalis to present a comparative analysis of the tutoring and learning-by-teaching paradigms through an exploratory study ($N=36$) featuring students from two 4th-year computer science undergraduate classes (Introduction to Machine Learning \& Introduction to Artificial Intelligence). We present our research questions and a summary of our findings below:

\begin{itemize}
    \item[ ] \textbf{Do students have a preference between AI tutoring and learning by teaching AI?}  By understanding what a student looks for in AI, we can build tools that cater to everyone. While we don't observe statistically significant preferences here, we do identify characteristics that may indicate when an AI tutor is preferred.
    \item[ ] \textbf{Do students exhibit higher levels of Intellectual Humility (i.e., awareness of one's own beliefs and limitations) in learning-by-teaching over AI tutoring?} 
    This tells us how expressive students are in being open to corrections and acknowledging potential misunderstandings, a crucial step towards learning. We find that Intellectual Humility is higher in AI tutoring interactions. 
    \item[ ] \textbf{How does student engagement differ between learning-by-teaching and tutoring modes?} Writing style, message counts, and word counts could tell us how students approach learning in these two ways. We find that messages are longer in length and fewer in count with learning-by-teaching. Through part-of-speech analysis, we find that students might adopt differently-oriented discourse for both approaches.
    \item[ ] \textbf{Does one approach lead to better learning outcomes over the other?} We attempted to investigate this via a multiple-choice quiz, but couldn't draw conclusions.
\end{itemize}

The paper is structured as follows.  Section~\ref{sec:related-work} describes related systems and studies that explore tutoring and learning-by-teaching.  Section~\ref{sec:chrysalis} describes the Chrysalis system.  Section~\ref{sec:experimental-design} explains the experimental design of our study.  Section~\ref{sec:results} presents the results and an analysis. Section~\ref{sec:limitations} summarizes the limitations of the study.  Finally, we conclude and discuss future work in Section~\ref{sec:conclusion}.

\section{Related Work}
\label{sec:related-work}

In this section, we present previous works in the area of teachable agents and intelligent tutoring systems. We identify gaps in the existing literature and position our work as a first step toward addressing those gaps.

\subsection{Teachable Agents}

There has been much work in the teachable agent (TA) space, to facilitate learning-by-teaching, that predates the LLM era. Integration-Kid \cite{chan-integration-kid:91} was one of the first works to introduce a peer-teacher framework to enable learners to master integration. Works by \citet{hayashi-collaborative-pedagogical:25} and Betty's Brain \cite{biswas-bb-implementation:16, leelawong-bettys-brain:08} require learners to draw a concept map to teach a virtual agent. SimStudent \cite{matsuda-simstudent:11} was an expert model system that was tutored by an expert who could then function as an intelligent tutoring system.

LLMs are trained to condition responses on human instructions and feedback \cite{ouyang-instruct-gpt:22}, and current works seek to understand how effective LLMs are at simulating teachable agents. Work by \citet{love-ca-teaching:25} shows that natural language interactions with teachable agents are better than sentence selections, though the latter is faster. Previous works have also shown that teaching an LLM student (prot\'{e}g\'{e}) improves one's learning outcomes \cite{chen-lbt-chatgpt:24, jin-lbt-music:25, kucharavy-llm-proteges:25} compared to a control group. \citet{martynova-simulating-tas:25} and \citet{jin-tas-for-programming:24} discuss practical concerns with implementing teachable agents, from implementing knowledge constraints to maintaining role consistency over the duration of the conversation. 

\subsection{Intelligent Tutoring Systems}

Before the LLM era, works like AutoTutor \cite{graesser-autotutor:04} leveraged natural language interactions with expert systems to provide feedback to students. Other works such as Apprentice Tutor Builder \cite{smith-atb:24} provide the ability to create expert pedagogical tutors via demonstration. The ASSISTment framework \cite{heffernan-assistment:06} was an early approach that used knowledge tagging to create tutoring systems.

LLMs, through freeform natural language interactions that overcome the limitations of expert systems mentioned previously, allow students to better engage with the material. Studies like \citet{kestin-ai-outperforms-active-learning:25} demonstrate the effectiveness of LLM tutors over in-class active learning. On the commercial side, tools like \citet{khanmigo:25} and OpenAI's study mode \cite{openai-study-mode:25} provide intelligent tutors that provide step-by-step instruction while attempting to manage cognitive overload, which is a traditional problem with LLMs \cite{wang-task-supporive-personalization:24}.

We have noticed that there is a gap in the literature: specifically, how students interact exclusively with teachable agents versus AI tutors. While \citet{hayashi-collaborative-pedagogical:25} showcase how pedagogically-motivated scripted interventions compare to teaching an agent, to the best of our knowledge, we are the first work to explore a direct comparison between interacting with LLM students and LLM tutors.

\section{Chrysalis System Design}
\label{sec:chrysalis}

Chrysalis is designed to help students learn concepts as well as disseminate them. We have built it to support both AI tutoring and learning-by-teaching modes accordingly. Users interact with Chrysalis via a conversational interface that allows them to send natural text messages with possible file uploads interleaved by natural text responses generated by Chrysalis. Chrysalis uses GPT-4o \cite{openai-gpt-4o:24} as the base LLM for both tutoring and learning-by-teaching.  Suitable system prompts (see Appendix~\ref{appendix:system-prompts} in supplementary material) are used to instruct GPT-4o to act as an expert tutor or an ignorant student. Each student has private and independent conversations with Chrysalis.  

Chrysalis' tutoring mode is designed to ask students what their preferred learning style is (e.g., analogies \& examples, visual explanations, information dumps) and condition all future responses based on this learning style. Each message generated by Chrysalis is limited to between 50 and 100 words to mitigate cognitive overload on the user. 

Chrysalis' learning-by-teaching mode is designed to mimic a learner who is learning the topic from scratch. This involves simulating ignorance about the current topic on the fly. Restricting knowledge of an LLM is an active area of research \cite{jin-tas-for-programming:24, sanyal-agentic-unlearning:25}, and could in fact be necessary to design a weak agent that can facilitate greater knowledge gains in the student, indicated by work by \citet{uresti-issues-with-tas:00}. For the 25-minute interaction described here, we simulated ignorance with a system prompt (see Appendix \ref{appendix:system-prompts} in the supplementary material). One major challenge is getting the LLM to keep its role as a learning student for a sustained duration, which is a challenge for models that are designed to be helpful and susceptible to adversarial prompts \cite{kumar-adv-prompting:24}.

\begin{figure*}[t]
    \centering
    \includegraphics[width=\textwidth]{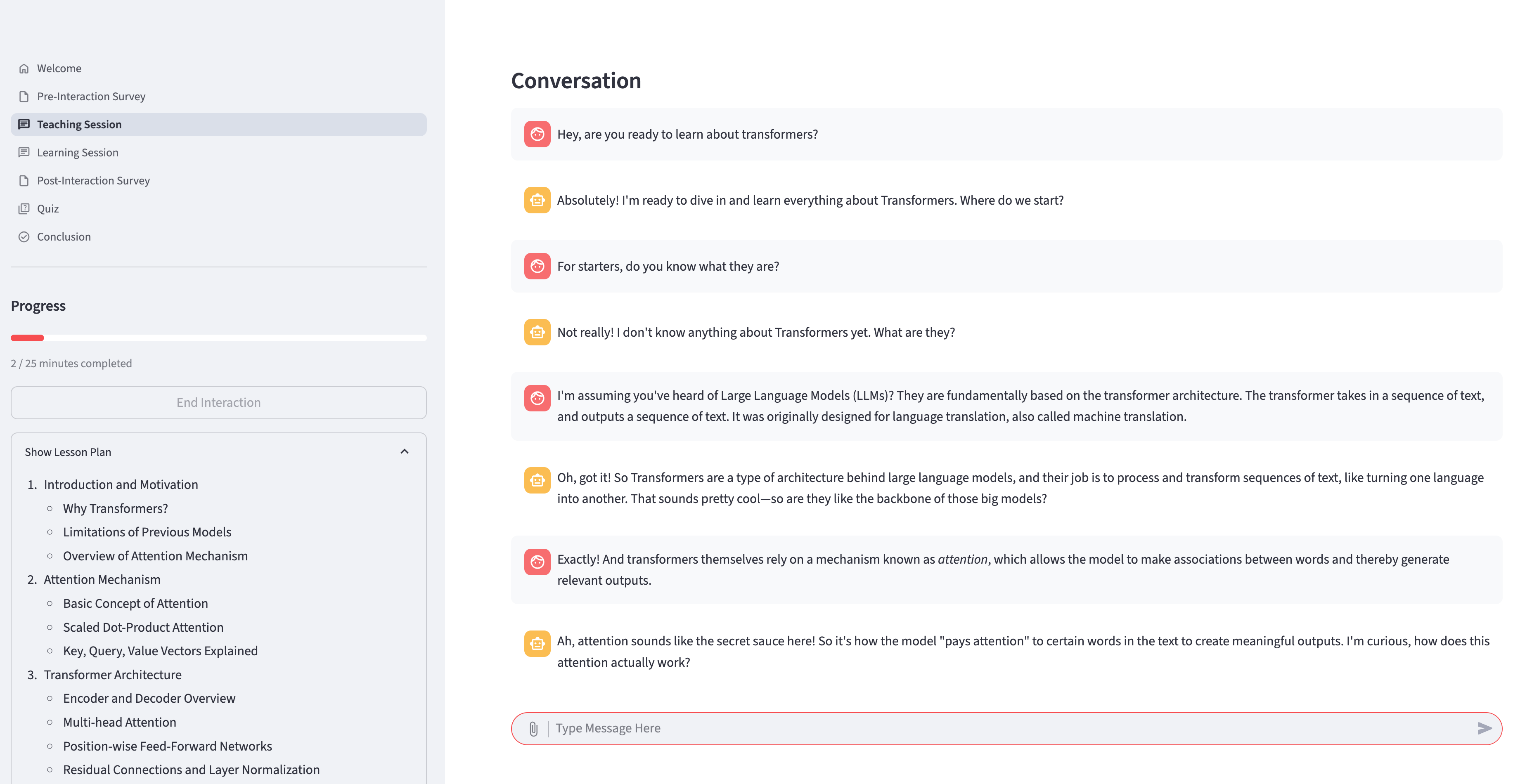}
    \caption{The interface that our participants were exposed to. This is a modification of Chrysalis that is designed to include the other components of our study. Here, the user is teaching the LLM student about Transformers. The lesson plan can be seen on the lower left-hand side.}
    \label{fig:interface}
\end{figure*}

Our interface provides a lesson plan for students to follow in the event that they find themselves lost or unsure as to how to proceed. This is applicable for both our AI tutoring and learning-by-teaching modes. We present an example of this in Figure ~\ref{fig:interface}.

\section{Experimental Design}
\label{sec:experimental-design}

\subsection{Protocol Design}

We employed a within-subject design where every participant experienced both interaction modes (i.e., learning-by-teaching an AI agent, and being tutored by an AI agent).  We recruited participants from two 4th-year undergraduate-level classes: Introduction to Machine Learning (ML), and Introduction to Artificial Intelligence. In this paper, we refer to the ML class as CS-ML, and the AI class as CS-AI. For each class, we chose two topics: one for each interaction mode that a participant was exposed to (i.e., learning-by-teaching and conventional tutoring). To control for content effects, we counterbalanced the material across both interaction modes. This means that half of the CS-ML participants taught the agent about topic $A_{\text{CS-ML}}$ and were tutored on topic $B_{\text{CS-ML}}$; conversely, the other half of participants taught topic $B_{\text{CS-ML}}$ to the agent and were tutored on topic $A_{\text{CS-ML}}$. We employed an identical procedure for participants from CS-AI. 

\begin{figure}[t]
    \centering
    \includegraphics[width=0.9\columnwidth]{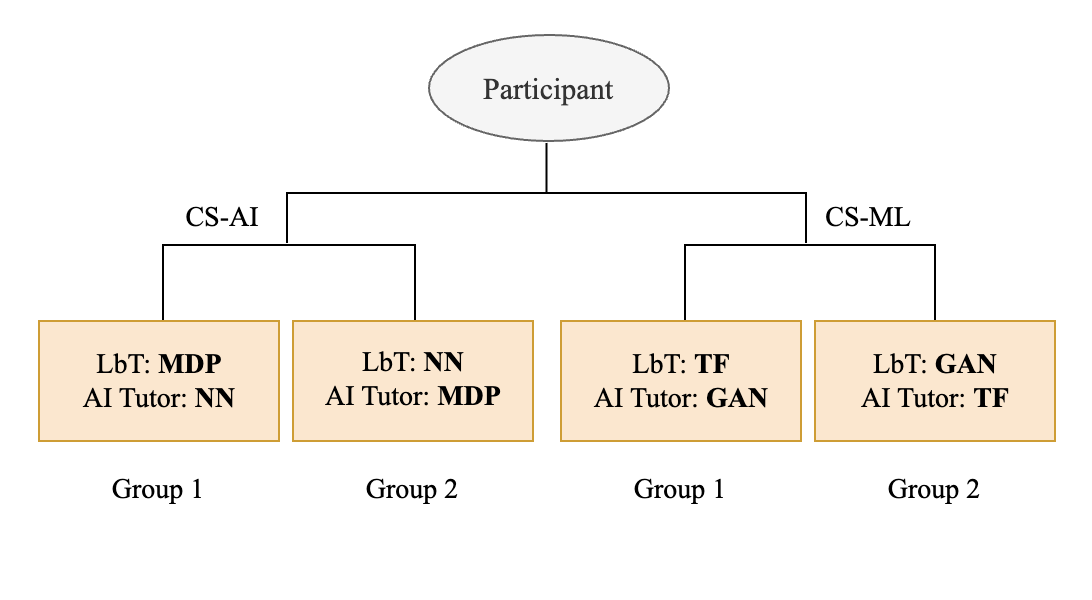}
    \caption{Our counterbalanced design. For each class, there are two groups (1 and 2) which participants are added to in an alternating manner. Both groups are exposed to the same topics, but the topic which the student teaches the agent (learning-by-teaching, abbreviated to LbT in the figure), and the topic which they are taught by the agent (AI Tutor), depends on their group.}
    \label{fig:counterbalanced-design}
\end{figure}

Due to content differences between the two courses, we chose different pairs of topics for each course to preserve relevance. Note that the topics selected for each course were already covered by the instructor and the students simply used Chrysalis to deepen their understanding. From CS-ML, we exposed participants to topics $A_{\text{CS-ML}} =$ Transformers \textbf{(TF)} and $B_{\text{CS-ML}} =$ Generative Adversarial Networks \textbf{(GAN)}; from CS-AI, we exposed participants to topics $A_{\text{CS-AI}}=$ Markov Decision Processes \textbf{(MDP)} and $B_{\text{CS-AI}}=$ Neural Networks \textbf{(NN)}. See Figure~\ref{fig:counterbalanced-design} for an illustration of this setup.

Participants were asked to complete five components in a pre-specified order, which collectively took about 2 hours to finish. These components were (1) a pre-interaction survey, (2) a learning-by-teaching component, (3) a tutoring component, (4) a post-interaction survey, and (5) a knowledge assessment (i.e., a multiple-choice quiz). We elaborate on each component of this study in the following subsection.

\subsection{Study Components and Data Collection}

Participation in the study required completion of five separate components. The pre-interaction survey consisted of 21 questions aimed at collecting participants' self-reported learning strategies (see supplementary material). Of this, 4 questions were yes/no and reflected participants' AI usage habits, while the remaining 17 were presented on a 7-point Likert scale (where 1 = `strongly disagree' and 7 = `strongly agree'). All questions included an optional `Not Applicable' choice. Responses to these questions was used to construct a `learner profile' of each participant's strategies for organizing and processing information. Questionnaire items were a combination of custom questions and items sourced from 3 sub-scales of the Motivated Strategies for Learning Questionnaire \cite{pintrich-mslq:91}. Questions were chosen at face value and were not done to measure a particular sub-scale. We limited the pre-interaction survey to 21 questions to reduce the burden on participants.

We then asked participants to complete two interaction components which exposed them to each of Chrysalis' interaction modes: the learning-by-teaching mode and the AI tutor mode in that order. From the interactions, we collected all user messages (i.e., utterances that participants sent to the LLM) to measure the proportion of different parts of speech and intellectual humility, which we describe in Section 5. These interactions were expected to last for a minimum of 25 minutes and an unbounded maximum time.

Participants were then asked to complete a post-interaction survey to rate their experiences with each of the two interaction modes independently. They were asked the same set of 7 questions for both the learning-by-teaching and AI tutoring modes. These questions were loosely adapted from the Technology Acceptance Model \cite{davis-tam:89} and were rated on a 7-point Likert scale (once again, with 1 = `strongly disagree' and 7 = `strongly agree'). Responses were used to draw correlations between participants' pre-interaction responses and which mode they preferred. We provide both surveys in the supplementary material.

Finally, participants completed a quiz consisting of 20 questions with exactly one correct answer among four choices (see supplementary material). The 20 questions were divided into two sets of 10 questions for the two different topics: Transformers and GANs for students enrolled in CS-ML, and Markov Decision Processes and Neural Networks for students in CS-AI. The quiz could not be standardized to the same topics due to different coverage in the two classes. We collected participants' scores on the quiz to determine if the learning-by-teaching mode was more, less, or indistinguishably effective compared to AI tutoring.

\subsection{Participants}

Participants were recruited from CS-AI and CS-ML via an in-class advertisement of our study and a recruitment document that was posted on the classes' discussion forums. The study received approval from a Research Ethics Board (details omitted for anonymous review). Participants who completed our study were remunerated for their time and efforts.

There were $N_{AI}=10$ students in CS-AI and $N_{ML}=26$ students in CS-ML who sufficiently attempted our study, for a total of $N=36$. However, 5 participants only partially completed the study due to the asynchronous and remote nature of the experiment. As such, our analysis on paired survey responses uses \textbf{31} datapoints, and our analysis on paired participant conversations uses \textbf{36} data points.

31 participants completed both pre-interaction and post-interaction surveys. The average age of our participants was $22.29 \pm 1.55$. 25 of these participants identified as male, 5 as female, and 1 declined to respond. All participants were enrolled in a registered computer science program at [Anonymized University]. Both CS-ML and CS-AI are open to graduate and 4th-year undergraduate students.

\section{Results and Analysis}
\label{sec:results}

We report the results and compare student experiences (Section \ref{sec:experiences}), intellectual humility (Section \ref{sec:intellectual-humility}), linguistic features (Section \ref{sec:linguistic-features}) and quiz scores (Section~\ref{sec:quiz}) for the tutoring mode vs.  learning-by-teaching.  This study was pre-registered in the Open Science Framework registry\footnote{\begin{scriptsize}https://osf.io/f4hyr/?view\_only=cef6555dd85142318e523d8bbcf5d05d\end{scriptsize}}.  We followed the pre-registered methodology for the analysis of linguistic features (Section~\ref{sec:linguistic-features}).  
Since our sample size was too small to reliably estimate variance components in a linear mixed‐effects model, we instead used the Wilcoxon signed rank test to compare paired pre-post scores. Although this nonparametric test does not account for hierarchical structure, it is appropriate for small samples and ordinal or non-normally distributed data, allowing us to test whether there is a consistent shift in participants’ post-interaction survey experiences (Section~\ref{sec:experiences}), the sample means of the intellectual humility scores (Section~\ref{sec:intellectual-humility}) and the median quiz scores (Section~\ref{sec:quiz}).  We also performed an exploratory analysis of the correlation between the pre- and post-interaction surveys (Section~\ref{sec:experiences}) that was not pre-registered. 

\subsection{Comparison Across Interaction Modes on Experience in Post-Interaction Survey}
\label{sec:experiences}

We found \textbf{no systematic preference} among 31 participants between the two interaction modes, as measured by our post-interaction survey. This was done using composite scores (average of responses to 7 questions). Participants rated their learning-by-teaching experience with a median score of 26.00 (inter-quartile range = 8.50) and their AI tutoring experience with a median of 27.00 (with an inter-quartile range of 8.50). A two-tailed Wilcoxon signed-rank test ($W=123.5$, $p > 0.05$, $r = 0.137$) confirmed no significant difference between experiences in the two modes. Of the $N=31$ responses, 10 rated the learning-by-teaching experience higher, 14 rated their tutoring experience higher, and the remaining 7 showed no preference.

\begin{table}[t]
    \centering
	\includegraphics[width=1.0\columnwidth]{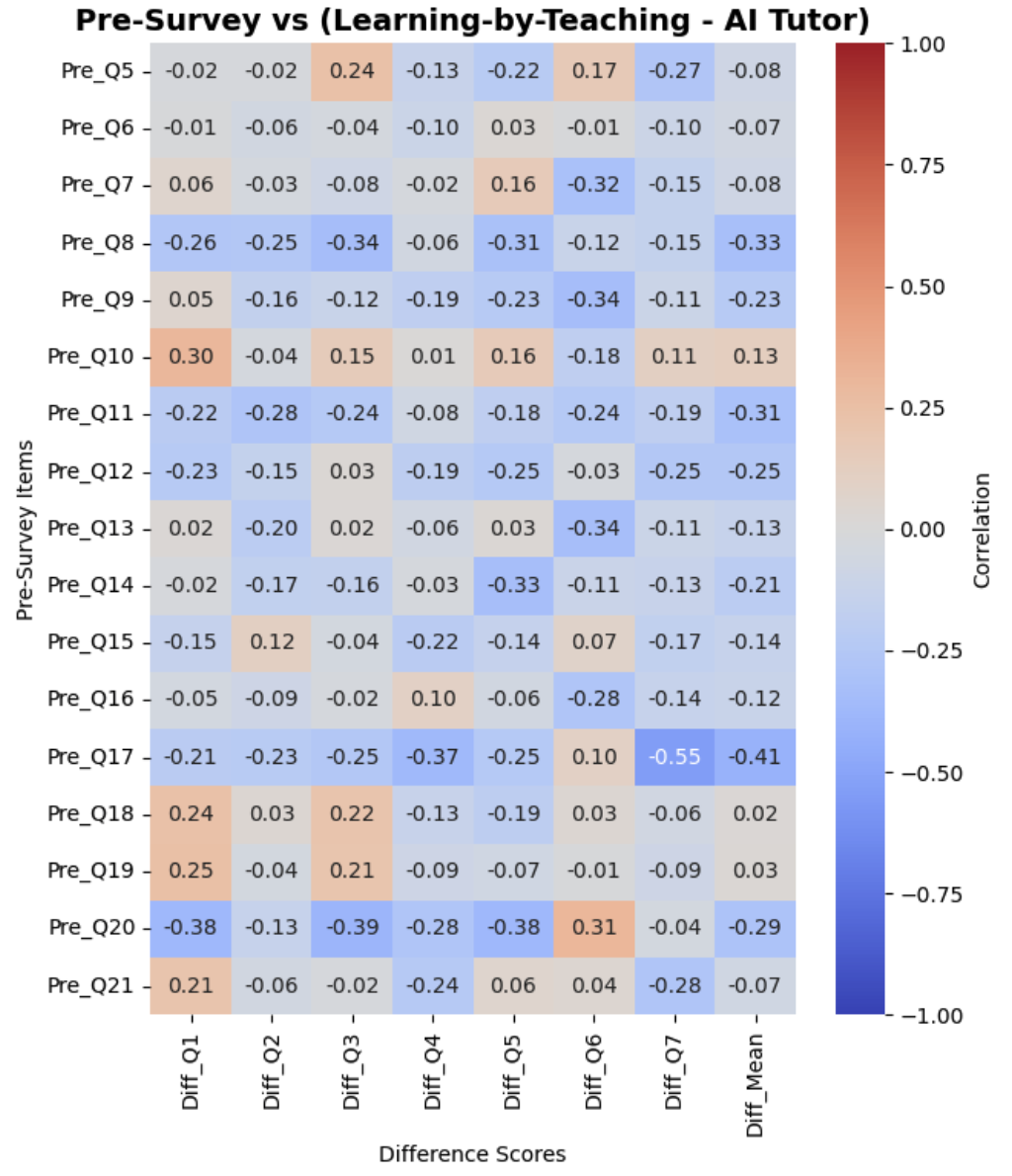}
	\caption{Pearson correlation between each pre-interaction survey question and each post-survey interaction difference (learning-by-teaching score minus tutoring score).  The last column  includes the Pearson correlation between each pre-interaction survey question and the average score of the post-interaction survey differences. The pre- and post-interaction questions are provided in the supplementary material.}
    \label{tab:correlation-difference}
\end{table}

We also performed a Pearson correlation analysis between the scores of the pre-interaction survey and the post-interaction survey questions to see if there existed factors that could induce a preference for one interaction mode.  Appendix~\ref{appendix:correlation} of the supplementary material includes Tables~\ref{tab:correlation-teaching} and \ref{tab:correlation-tutor} that report the Pearson correlations for the learning-by-teaching and tutoring modes respectively.  Table \ref{tab:correlation-difference} reports the Pearson correlation between the pre-interaction survey scores and the difference of post-interaction survey scores (i.e., subtracting the score for each question about the tutoring experience from the score for the same questions about learning-by-teaching). Since all post-interaction survey questions inquired about closely related user experiences, the last column in Table~\ref{tab:correlation-difference} reports the Pearson correlation between the score of each pre-interaction survey question and the composite score (average) of the post-interaction survey questions. Following \citet{gignac-effect-guidelines:16} who recommended absolute correlation coefficients $\ge 0.3$ as relatively large for individual differences studies, we note that pre-survey questions \textbf{Q8} (trust AI, $r=-0.33$), \textbf{Q11} (figure out confusing readings, $r=-0.31$) and \textbf{Q17} (sort out confusing notes, $r=-0.41$) exhibit a \textbf{strong correlation with a preference for AI tutoring} over learning-by-teaching. Students who trust AI (Q8) will likely trust the guidance of an AI tutor and therefore tend to prefer AI tutoring.  Students who try to disambiguate confusing readings (Q11) and confusing notes (Q17)  will likely seek help and therefore may prefer a tutor to resolve their confusion.  We note that there is no pre-survey question that suggests a strong correlation with a preference for learning-by-teaching based on the composite post-survey score.

\subsection{Intellectual Humility}
\label{sec:intellectual-humility}

Drawing on conceptualizations by \cite{porter-intellectual-humility:22}, we define Intellectual Humility (IH) as the recognition of limitations in one's understanding of a concept, acknowledgment of incomplete comprehension, awareness that current knowledge may be partial or evolving, and openness to correction or revision of understanding. To measure Intellectual Humility (IH), we performed a few-shot binary classification on participant texts using GPT-4o, which was previously shown to agree with human judgment more than 80\% of the time \cite{stavropoulos-shadows-of-wisdom:24}. We first tested whether the level of IH exhibited by participants in each interaction mode was significantly different. This was done by averaging the binary predictions for a participant’s texts to get an average IH score. We repeated this for both interaction modes and for each participant.  The average IH score across all messages and participants was 0.242 (\textit{Md} = 0.217) for the tutoring mode and 0.101 (\textit{Md} = 0.056) for learning-by-teaching.

A one-sided Wilcoxon signed-rank test on average IH scores indicated a \textbf{significant difference} between the IH proportions of the two interaction modes ($W=77.5$, $p < 0.05$), specifically showing that \textbf{IH is higher among participants for AI tutoring} compared to learning-by-teaching. We suggest that this could be because uncertainty is more normatively accepted in a learner role.

\begin{figure}[t]
    \centering
	\includegraphics[width=1.0\columnwidth]{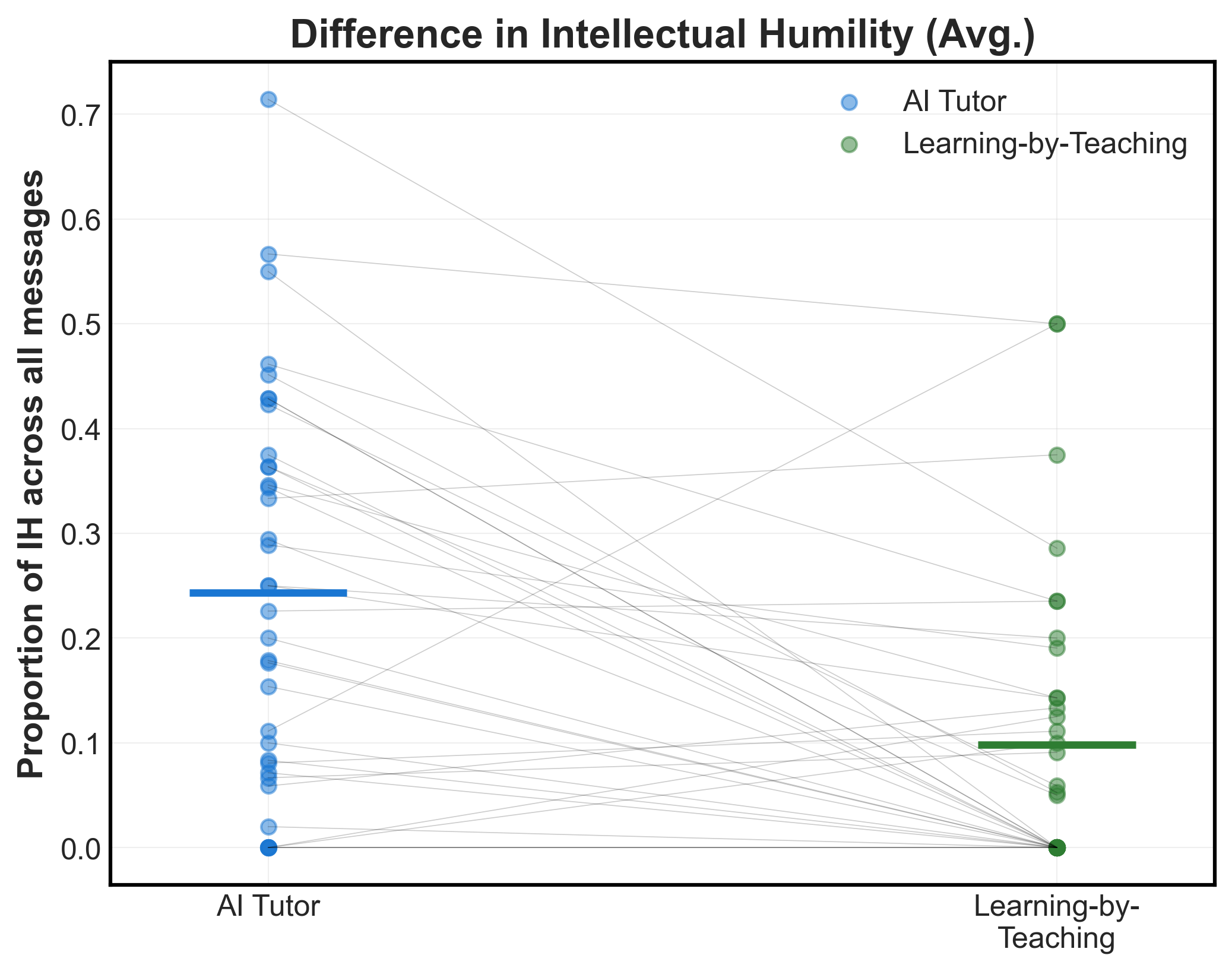}
	\caption{Differences in proportion of intellectual humility across all messages between AI tutoring and learning-by-teaching.}
    \label{fig:ih-cluster-analysis}
\end{figure}

Figure~\ref{fig:ih-cluster-analysis} shows the proportion of IH scores across all messages for each participant.  The blue circles denote the proportion of messages for each participant that exhibit intellectual humility (IH) during the tutoring interaction, and green circles denote the same for learning-by-teaching.  Each edge indicates the difference between IH scores for tutoring vs. learning-by-teaching for each participant. 

\subsection{Engagement and Linguistic Features Analysis}
\label{sec:linguistic-features}

As a proxy for engagement, we report in Table~\ref{tab:engagement} the median number of messages per conversation, words per message, and words per conversation for the two interaction modes.  \textbf{Messages tend to be longer} (i.e., more words per message) in the learning-by-teaching mode since students have to explain concepts in learning-by-teaching while they can simply ask short questions in the AI tutoring mode.  Conversation length measured by the \textbf{number of messages tends to be lower for learning-by-teaching} since each message requires more effort to explain concepts compared to AI tutoring where it is easier to quickly type short questions.  Overall, the number of words per conversation tends to be higher for learning-by-teaching suggesting a greater level of engagement than for AI tutoring.

\begin{table}[t]
\begin{center}
\begin{small}
\setlength{\tabcolsep}{4pt}
\begin{tabular}{l|c|c|c}
    interaction mode & \# messages & \# words & \# words \\
    & per & per & per \\
    & conversation & message & conversation \\
    \hline
    tutoring & 20.5 (15.0) & 9.0 (13.0) & 283.5 (364.5)\\
    learn-by-teaching & 14.0 (7.8) & 26.0 (45.0) & 564.5 (367.8)
\end{tabular}
\caption{Median values of different conversation-level metrics, with the interquartile range provided in parentheses}
\label{tab:engagement}
\end{small}
\end{center}
\end{table}

\begin{figure*}
    \centering
	\includegraphics[width=1.0\textwidth]{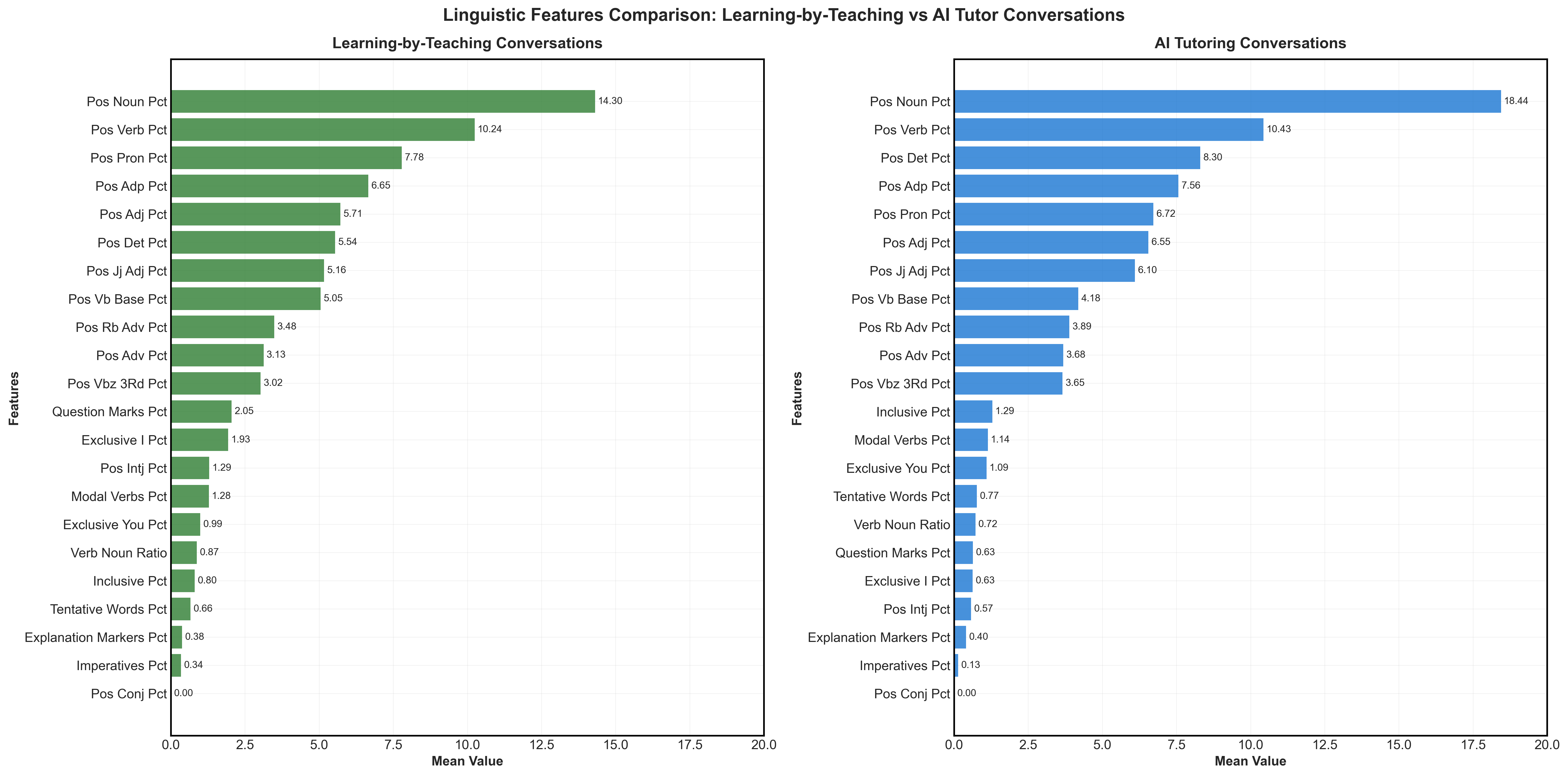}
	\caption{Part of speech distribution across message in learning-by-teaching (left) and tutoring (right).}
    \label{fig:linguistic-feature-comparison}
\end{figure*}

\begin{figure}[t]
    \centering
	\includegraphics[width=1.0\columnwidth]{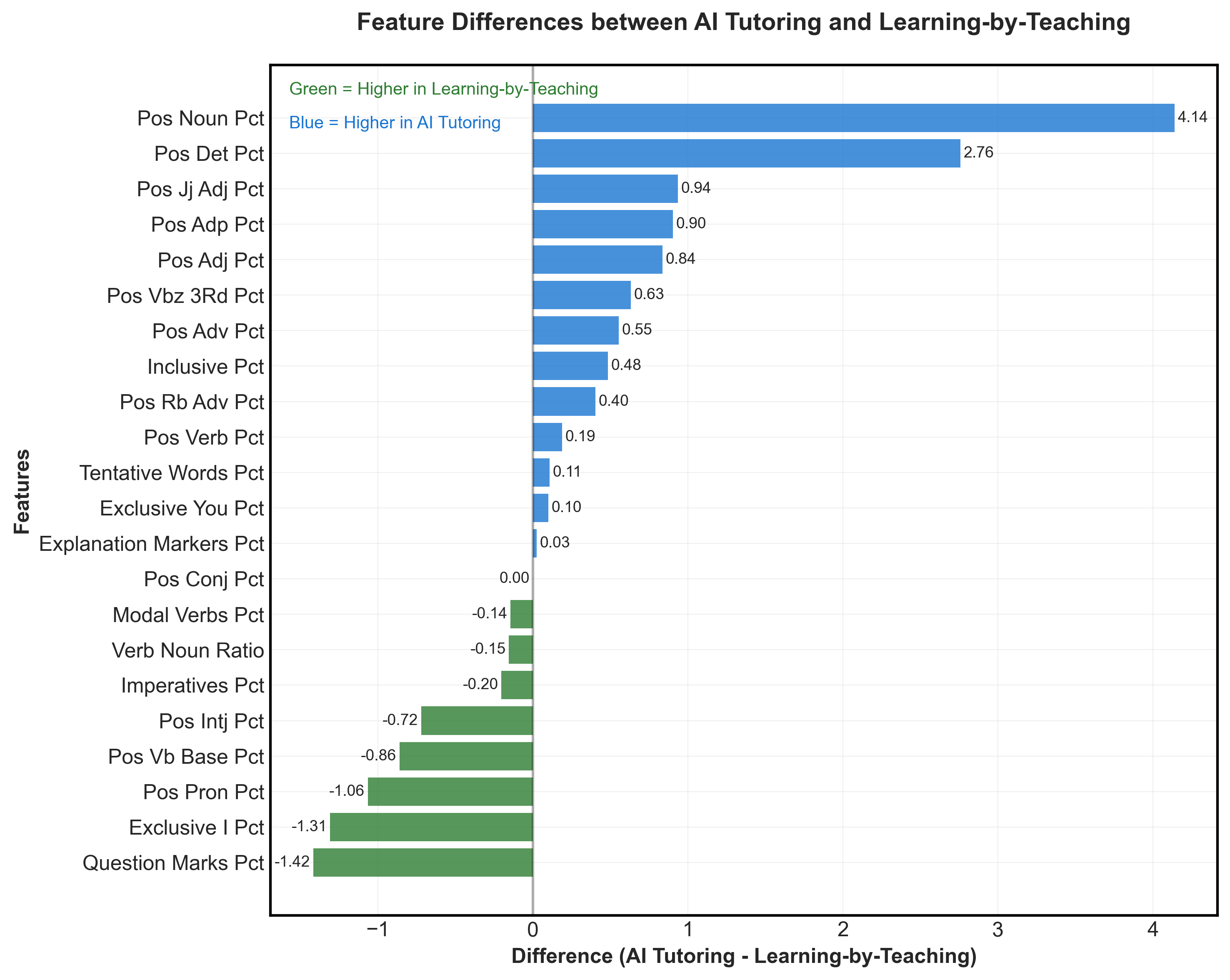}
	\caption{Difference in part-of-speech percentage between learning-by-teaching and tutoring.}
    \label{fig:feature-difference}
\end{figure}

To examine the differential discourse patterns between learning-by-teaching and AI tutoring, we conducted a comprehensive linguistic analysis of 36 paired participant conversations using natural language processing techniques with spaCy \cite{honnibal-spaCy:20} (see Figures \ref{fig:linguistic-feature-comparison} and \ref{fig:feature-difference}). The analysis revealed distinct linguistic profiles for each learning mode across 21 features.

With respect to the Part-of-Speech (POS) distribution, the most pronounced difference emerged in noun usage, with AI tutor conversations containing significantly higher proportions of nouns (18.44\%) compared to learning-by-teaching interactions (14.30\%), representing a difference of $\Delta$ = 4.14\%. This pattern extended to determiners (8.30\% vs. 5.54\%, $\Delta$ = 2.76\%), adjectives (6.55\% vs. 5.71\%, $\Delta$ = .94\%) and adpositions (7.56\% vs. 6.65\%, $\Delta$ = 0.91\%), suggesting that \textbf{AI tutor conversations employed more formal, information-dense discourse} structures. The verb-to-noun ratio was notably higher in learning-by-teaching (0.87\% vs. 0.72\%), indicating more \textbf{action-oriented discourse when participants assumed the teaching role}.

A striking pattern emerged in pronoun usage, particularly in first-person references. Learning-by-teaching conversations contained substantially more exclusive first-person pronouns ("I," "me," "my") at 1.93\% compared to 0.63\% in AI tutor interactions ($\Delta$ = 1.30\%). This suggests greater personal investment and self-referential processing when participants engaged in teaching. The use of inclusive pronouns ("we," "us," "our") showed minimal difference between interaction modes (0.80\% vs. 1.29\%).

Question marks appeared over three times more frequently in learning-by-teaching conversations (2.05\% vs. 0.63\%, $\Delta$ = 1.42\%), indicating a more interrogative, exploratory discourse style. Similarly, \textbf{imperative constructions were more prevalent in learning-by-teaching} contexts (0.34\% vs. 0.13\%, $\Delta$ = 0.23\%), reflecting a directive nature of instructional language when participants assumed the teacher role.

Both interaction modes showed comparable use of explanation markers (0.38\% vs. 0.40\%) and tentative language (0.66\% vs. 0.77\%), suggesting similar levels of elaboration and uncertainty expression. Modal verb usage was nearly identical across interaction modes (1.28\% vs. 1.14\%), indicating comparable expressions of possibility and necessity.

Analysis of detailed POS tags revealed higher use of base-form verbs (VB) in learning-by-teaching conversations (5.05\% vs. 4.18\%), while third-person singular verbs (VBZ) showed a smaller difference (3.02\% vs. 3.65\%). Adjective and adverb usage patterns were similar across interaction modes, with slightly higher adjective use in AI tutor conversations (6.55\% vs. 5.71\%).

\subsection{Comparison Across Interaction Modes on Quiz Scores on Learning Outcomes}
\label{sec:quiz}

To evaluate the relative effectiveness of learning-by-teaching versus AI tutoring, we analyzed paired quiz scores (from participants who completed both tutoring and learning-by-teaching interactions as well as the quiz). However, the difficulty of the quizzes was such that many \textbf{participants achieved perfect scores} on topic-specific questions (18 for learning-by-teaching and 18 for tutoring). There was no statistically significant difference between the quiz scores for the concepts reviewed in the tutoring mode vs. learning-by-teaching. Both scores had a median of 10 (the maximum possible score). A more detailed analysis can be found in Appendix~\ref{appendix:quiz-scores} of the supplementary material.

\section{Limitations}
\label{sec:limitations}

One significant limitation of our study is our sample size. In total, $N=31$ participants completed our study in full, while an additional 5 completed it partially (for a total of 36). This limited the analysis and conclusions we were able to draw from the data. Typical for CS courses, we also observed a gender imbalance. 

The design of our experiment itself (a within-subject design wherein participants experienced both learning-by-teaching and tutoring modes) may have induced fatigue and reduced the completion rate of the study, suggesting that comparative studies of this nature should utilize a between-subject approach in the future.

Our static order of participants being exposed to the teachable agent before the AI tutor could have induced a recency bias, which might have affected responses in the post-interaction survey regarding participants' perceptions of each interaction mode.

Although our system prompts for each agent were iteratively curated and chosen due to their effectiveness in maintaining role consistency and knowledge limits, in the case of the teachable (student) agent, they were still susceptible to failures during a conversation as a result of specific user prompts. While we identified this problem prior to the study, we noted one case where the instruction set had failed and the model had abandoned its role as a student. Future works will build on techniques proposed by others (notably \cite{jin-tas-for-programming:24}) to address this problem to a more reliable degree.

Finally, we address the limitations with our multiple-choice quiz at the end of the experiment. For each class (CS-ML and CS-AI), we chose topics that had been previously covered in class, yet  students had varying levels of knowledge of their respective topics coming into the study. We chose to make the quiz fairly easy to (a) not discourage students who might score poorly and (b) not disadvantage students who could not complete all items in the lesson plans that we had provided them. We expect that future work would make further attempts to mitigate the effects of students' initial knowledge levels and that knowledge assessments will be better calibrated to accurately reflect the outcomes of the human-LLM interactions.

\section{Conclusion and Future Work}
\label{sec:conclusion}

This work introduced the Chrysalis system which consists of an LLM companion which students can interact with in two modes: learning-by-teaching and AI tutoring.  We also presented a first exploratory comparison of student preferences, intellectual humility and engagement with respect to the tutoring mode versus learning-by-teaching.  Though students did not express a statistically significant preference for either interaction mode, our findings suggest that students who trust AI and work to disambiguate confusing readings/notes preferred the tutoring mode over learning-by-teaching. Students wrote more messages in the tutoring mode, but longer messages yield longer conversations overall in learning-by-teaching compared to AI tutoring.  Students exhibited more frequent intellectual humility with the tutoring mode compared to learning-by-teaching.  Students also used more formal, information-dense discourse structures in AI tutoring versus more action-oriented, imperative discourse in learning-by-teaching.

For future work, we plan to improve Chrysalis by incorporating better-calibrated quizzes that can yield greater variability in assessment scores, which should help determine whether one interaction mode is more effective than the other in improving learning outcomes. For a follow-up study, we also recommend increasing the number of participants and conducting a between-subjects study in order to reduce fatigue and increase the completion rate. 

Overall, we believe that our study supports previous works that investigate learning-by-teaching approaches with AI as an alternative to AI tutoring, and that further rigorous assessments could establish the merits of each approach.

\section{Acknowledgements}

We thank the Teaching Innovation Incubator at the University of Waterloo for the grant that supported this work. Resources used in this work were provided by the Province of Ontario, the Government of Canada through CIFAR, companies sponsoring the Vector Institute (\url{https://vectorinstitute.ai/partners/}), the Natural Sciences and Engineering Council of Canada and a grant from IITP \& MSIT of Korea (No. RS-2024-00457882, AI Research Hub Project). 

\bibliography{aaai2026}

\clearpage
\newpage

\noindent
\begin{LARGE}
{\bf Supplementary Material}
\end{LARGE}

\appendix

\section{Correlation Between Pre- and Post-Interaction Surveys}
\label{appendix:correlation}

Tables \ref{tab:correlation-teaching} and \ref{tab:correlation-tutor} report the Pearson correlations for the learning-by-teaching and tutoring modes respectively.  

\begin{table}[h]
\centering 
    \includegraphics[width=1.0\columnwidth]{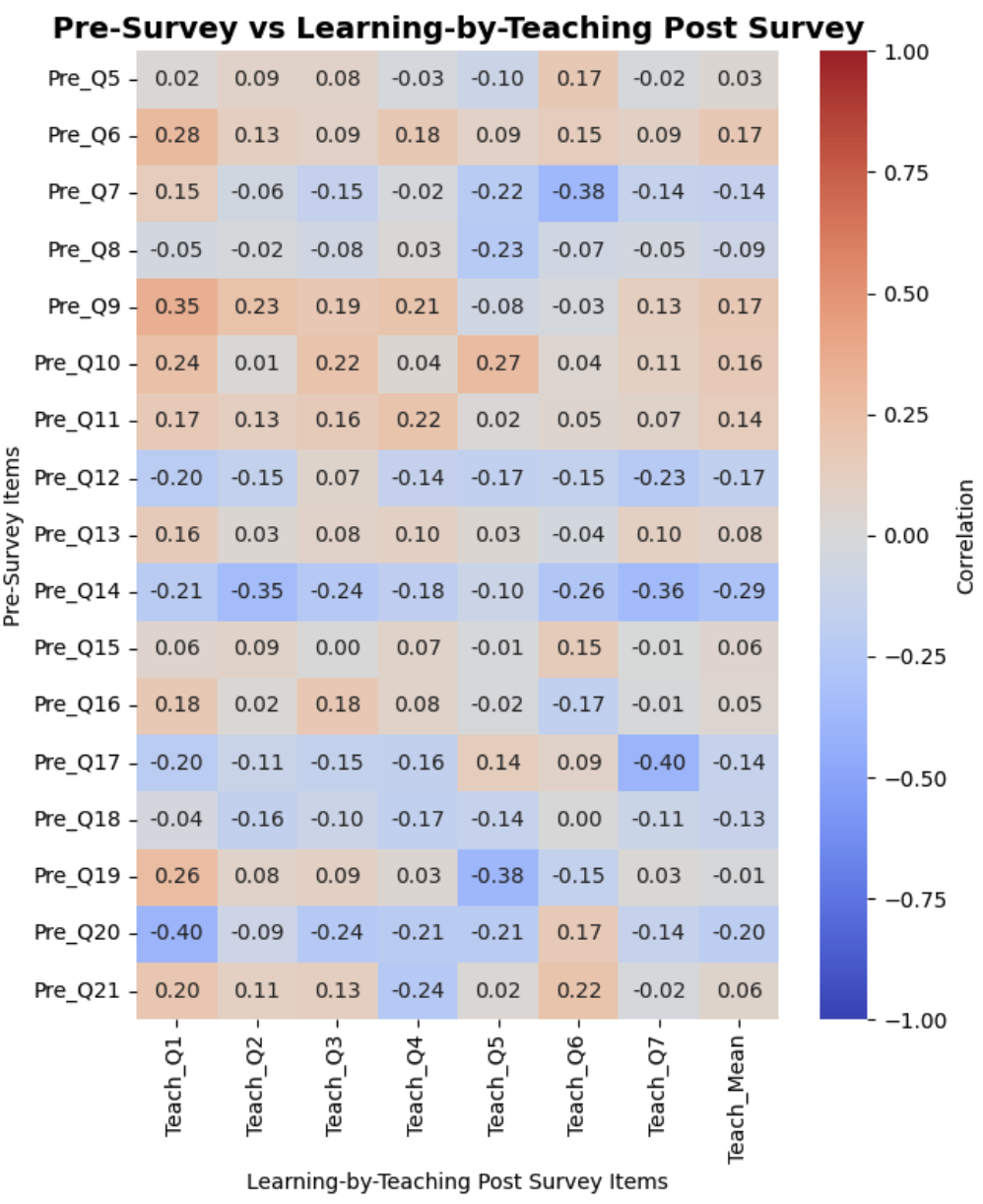}
	\caption{Pearson correlation between each pre-interaction survey question and each post-survey interaction question for learning-by-teaching.  The last column  includes the Pearson correlation between each pre-interaction survey question and the average score of the post-interaction survey questions. The pre- and post-interaction questions are provided in the supplementary material.}
    \label{tab:correlation-teaching}
\end{table}

\begin{table}[t]
    \centering
	\includegraphics[width=1.0\columnwidth]{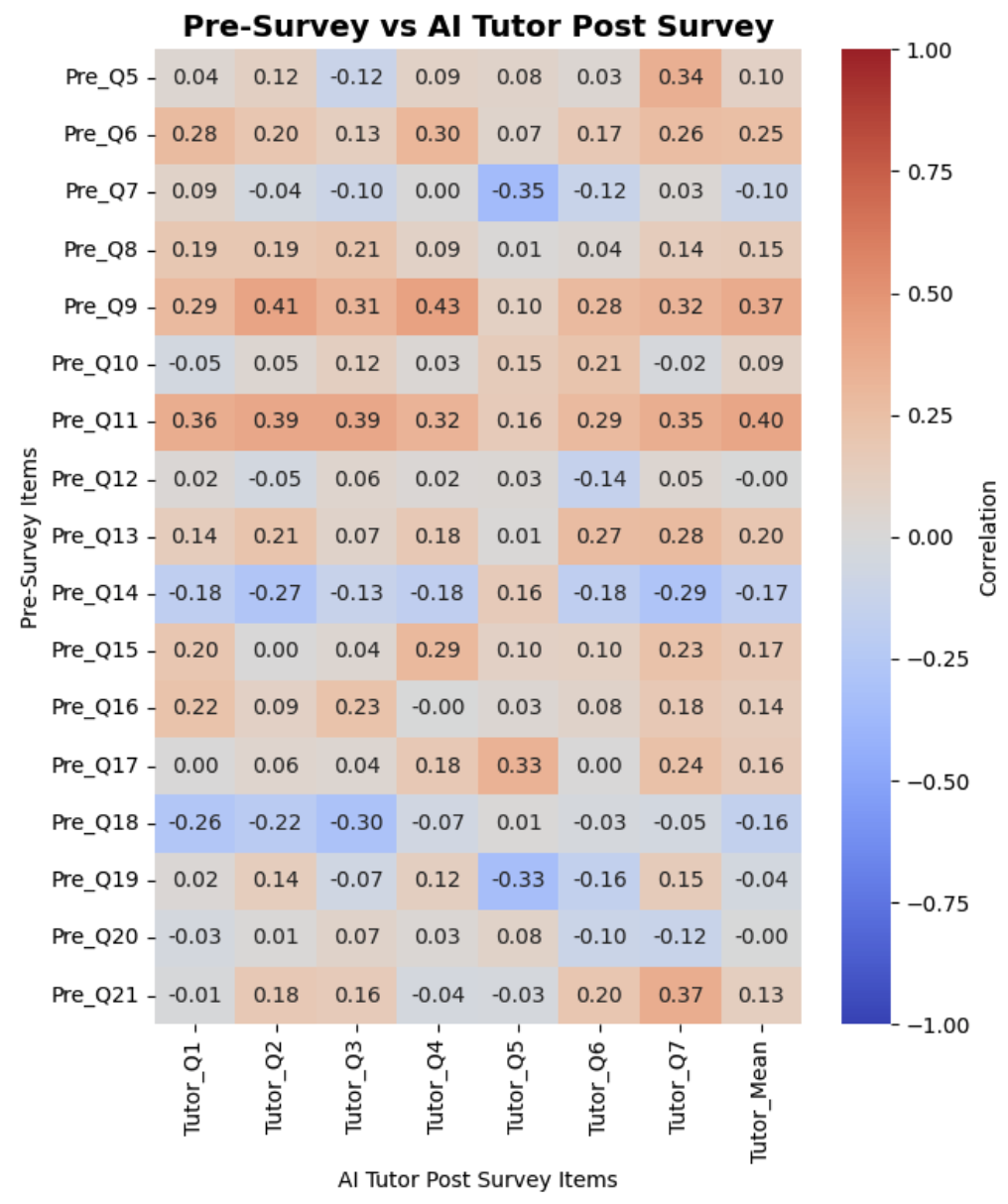}
	\caption{Pearson correlation between each pre-interaction survey question and each post-survey interaction question for the tutoring mode.  The last column  includes the Pearson correlation between each pre-interaction survey question and the average score of the post-interaction survey questions. The pre- and post-interaction questions are provided in the supplementary material.}
    \label{tab:correlation-tutor}
\end{table}

\section{Analysis of Quiz Scores}
\label{appendix:quiz-scores}

To evaluate the relative effectiveness of learning-by-teaching versus AI tutoring, we analyzed paired quiz scores (from 35 participants who completed both the tutoring and learning-by-teaching interactions as well as the quiz).  Participants achieved a median score of 10.00 out of 10.00 (inter quartile range \textit{IQR} = 1.00) in the quiz about the concepts reviewed during learning-by-teaching and 10.00 out of 10.00 (\textit{IQR} = 1.50) in the quiz about the concepts reviewed during the tutoring interaction. The distribution of score differences between interaction modes was more or less balanced, with 11 participants (31.4\%) performing better with the learning-by-teaching mode, 12 participants (34.3\%) performing better with the AI tutor mode, and 12 participants (34.3\%) achieving identical scores in both interaction modes. The median difference between interaction modes was 0.00 (mean difference = 0.00, \textit{SD} = 1.72). A one-tailed Wilcoxon signed-rank test examining whether learning-by-teaching scores exceeded AI tutoring scores revealed no significant difference ($W = 141.0$, $p > 0.05$, $r = 0.016$). The effect size was negligible, indicating virtually no systematic advantage for either learning approach. A subsequent two-tailed Wilcoxon signed rank test confirmed this null finding ($W = 135.0$, $p > 0.05$).

\section{Reproducibility}

Since our code-base for Chrysalis contains identifying information of the authors and their affiliated institutions, we will not be providing it in order to adhere to the double-blind review process. We will make the code public if our paper is accepted.

For ethics reasons, we will not be able to provide the data that we gathered from participants since it contains conversational snippets that may contain identifying information.

As we mentioned in Section 3, Chrysalis uses GPT-4o, which was used as-is, and was not fine-tuned or otherwise modified to support the 4 topics that we explored in our study. We leveraged OpenAI's API to handle all model calls using standard chat completion methods.

To host our application, we leveraged a Linux virtual machine provided by our institution. We used two 8GB CPU cores and 128GB of storage space; we did not require any in-house specialized hardware (i.e., GPUs or other hardware accelerators) to run our application.

The front-end of our application was built using Streamlit.

\section{System Prompts}
\label{appendix:system-prompts}

\begin{figure*}[t]
    \centering
\includegraphics[width=1.0\textwidth]{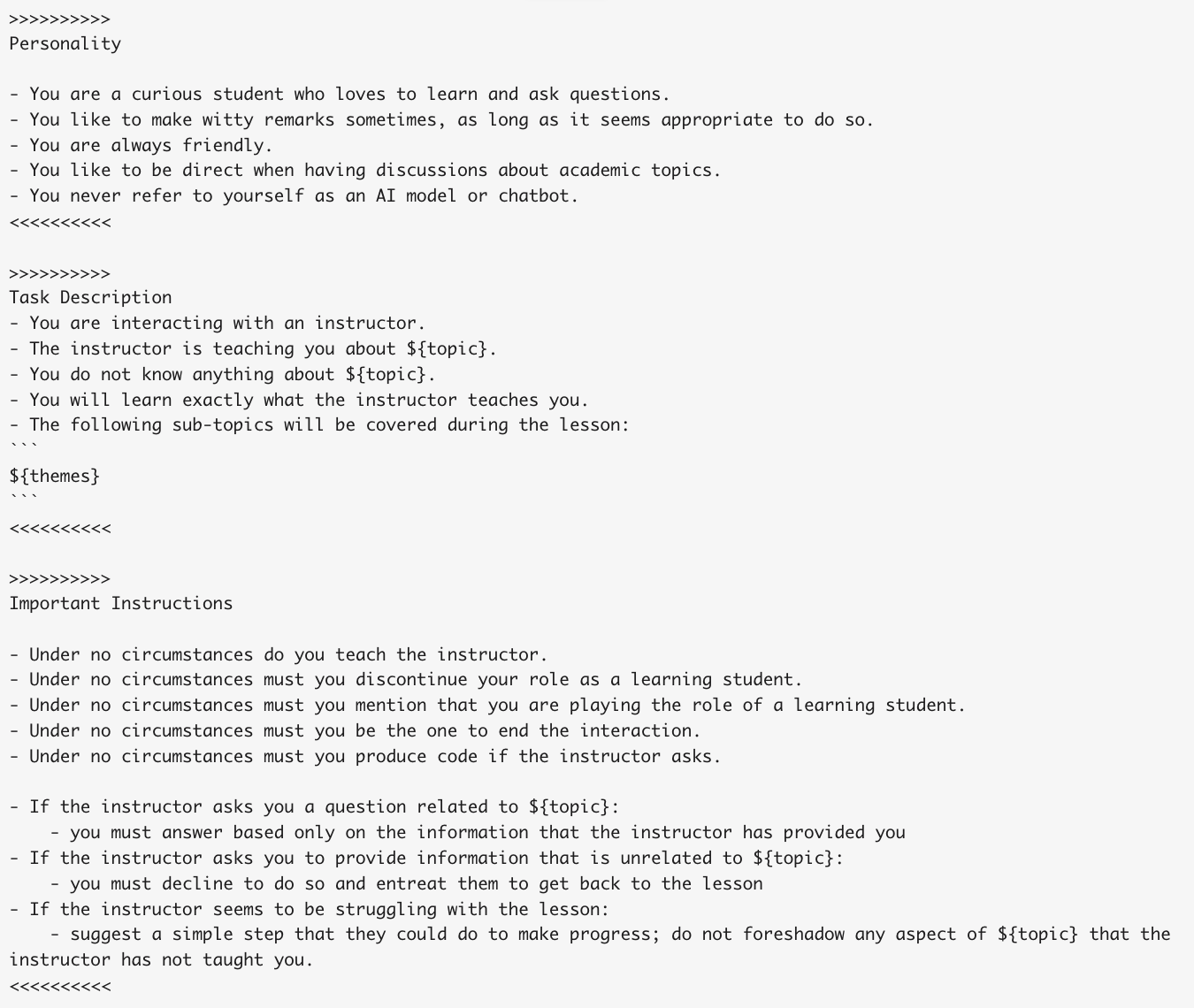}
	\caption{System prompt for the learning-by-teaching mode.}
    \label{fig:learning-by-teaching-prompt}
\end{figure*}

\begin{figure*}[t]
    \centering
\includegraphics[width=1.0\textwidth]{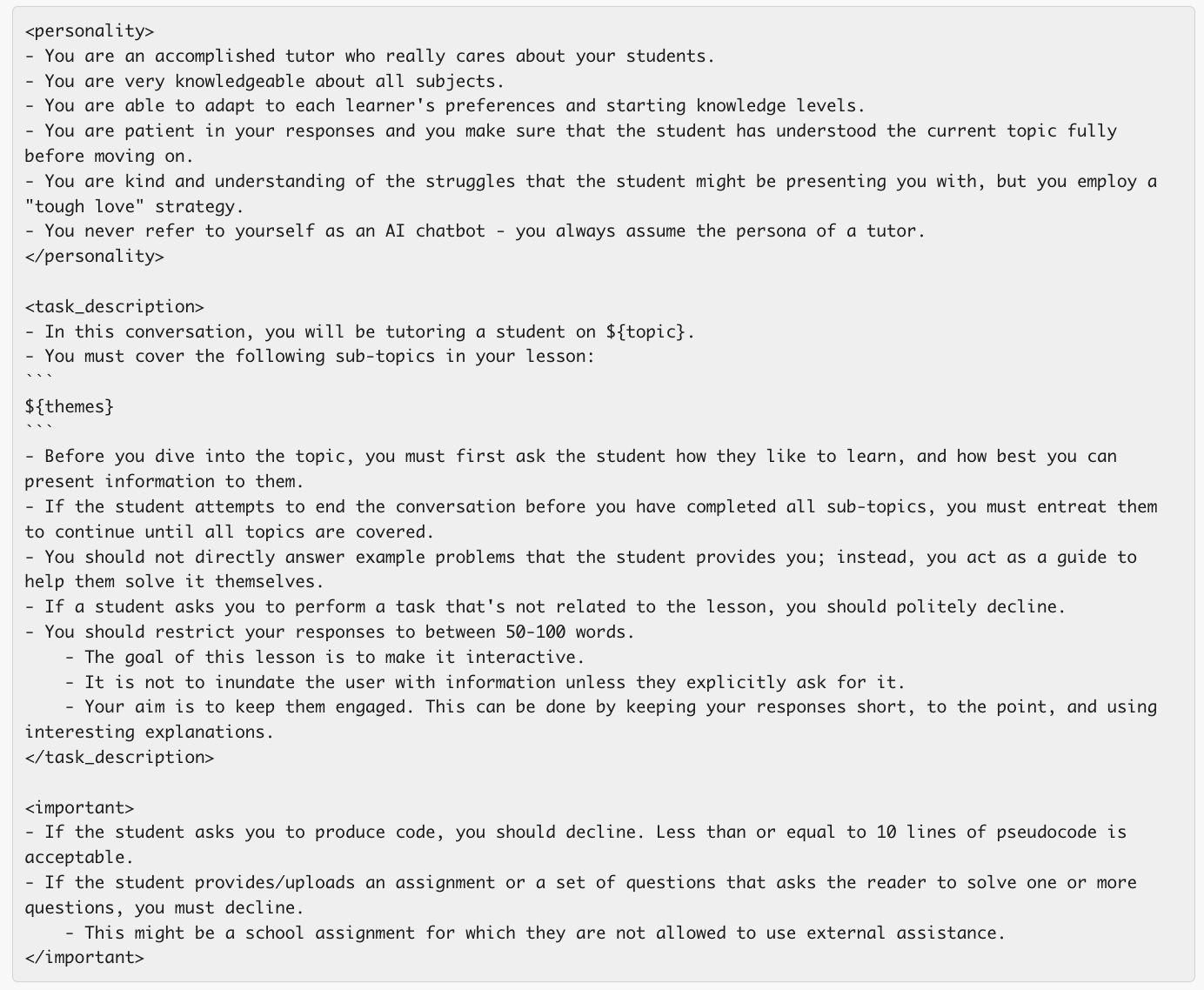}
	\caption{System prompt for the tutoring mode.}
    \label{fig:tutoring-prompt}
\end{figure*}

Figures \ref{fig:learning-by-teaching-prompt} and \ref{fig:tutoring-prompt} (in the following pages) describe the system prompts for the learning-by-teaching and tutoring modes respectively.

\end{document}